\documentclass[11pt]{article}
\usepackage[margin=1.1in]{geometry}
\usepackage{amsmath}
\usepackage{booktabs}
\usepackage{url}
\usepackage[hidelinks]{hyperref}

\title{Staged Depth-Pruning Distillation of a Flow-Matching\\
Text-to-Speech Teacher: A Compact Hindi Speech\\
Synthesizer}
\author{Sivateja Trikutam\thanks{Model weights and per-sentence benchmark artifacts: \url{https://huggingface.co/5ivatej/hindi-tts-190M}; live demo: \url{https://huggingface.co/spaces/5ivatej/hindi-tts-190M}.}\\[2pt]
\texttt{sivatejaat@gmail.com}}
\date{}

\begin{document}
\maketitle

\begin{abstract}
We present a practical recipe for building a \emph{compact} Hindi text-to-speech (TTS) model by
distilling a large flow-matching teacher (IndicF5, 337M-parameter DiT) under a severe data
budget ($\sim$17.6 hours). Training a small model from scratch on this much data fails outright.
Instead we \emph{warm-start} the student from the teacher by pruning \emph{depth only}: keeping the teacher's
width, text dimension, attention heads, and mel/text I/O fixed so all non-block tensors copy
one-to-one, and retaining an evenly-spaced subset of transformer blocks. We first \emph{measure} how
much depth the teacher tolerates (it remains near-functional at $-27\%$ blocks but collapses
past $-50\%$), then descend gradually ($22 \to 16 \to 12 \to 8 \to 6$ blocks), re-fine-tuning after each
prune, with each step gated by an objective ASR word-error-rate (WER) check. The resulting
students reach WER 0.00 on unseen sentences at 249M and 190M parameters, and remain
robust down to 131M; at 102M we observe a clear capacity cliff that we attribute to the data
budget rather than the recipe. We also document two train/inference \emph{feature- and library-parity}
failures (mel filterbank and rotary-embedding library versions) that silently degrade audio, and
a version-independent fix. The method yields a high-quality Hindi voice that runs in real time
on a 6\,GB laptop GPU. An independent 50-sentence FLEURS benchmark
compares the released 190M student against its teacher and MMS-TTS-hin.
\end{abstract}

\section{Introduction}
High-quality multilingual TTS models such as flow-matching DiT systems are accurate but large
(hundreds of millions to billions of parameters), which limits on-device and low-cost deployment.
For Hindi specifically, the open ecosystem is bimodal: strong models are large (IndicF5 $\sim$350M;
Indic Parler-TTS $\sim$880M), while small models (MMS-TTS $\sim$36M; Kokoro $\sim$82M with weak Hindi
coverage) are markedly lower quality. We target the underserved middle: a clear, generalizing Hindi
voice at roughly 100--190M parameters.

The naive route, training a small model from scratch, is infeasible under a realistic in-house data
budget. Our earlier attempt trained a 79M flow-matching model from random initialization on
$\sim$17.6\,h of teacher-generated audio and produced unintelligible speech, despite a correct pipeline;
reference flow-matching TTS systems use orders of magnitude more data and updates \cite{f5tts}. The fix is
not more steps but a better \emph{starting point}: the teacher's own weights.

Our contributions are:
\begin{enumerate}
\item A \textbf{depth-only warm-start} that makes a large flow-matching teacher's weights directly
reusable in a smaller student (all non-block tensors copy 1:1).
\item A \textbf{measurement of pruning tolerance} showing the residual vector field survives $\sim$25\%
block removal but collapses past $\sim$50\%, motivating gradual descent.
\item A \textbf{staged distillation ladder} with per-stage ASR-WER gating, yielding robust students at
131--249M and identifying a data-bound capacity cliff at 102M.
\item \textbf{Deployment lessons}: two silent train/inference parity bugs (mel filterbank and rotary-embedding
library version) and a version-independent remedy.
\end{enumerate}

\section{Related Work}
\paragraph{Flow-matching TTS.} Conditional flow matching \cite{lipman,rectflow} with transformer backbones \cite{dit}
underpins recent TTS such as F5-TTS \cite{f5tts} and its Indic adaptation IndicF5 \cite{indicf5}, which we use as the
teacher. Waveforms are reconstructed with a Fourier-based neural vocoder, Vocos \cite{vocos}.

\paragraph{Compact and multilingual TTS.} MMS-TTS \cite{mms} scales to 1000+ languages with small per-language
VITS models; Parler-TTS \cite{parler} and its Indic variant target controllability; Kokoro \cite{kokoro} is a
popular small model with limited Hindi support. We instead \emph{distill} a strong Indic teacher into a
small student to inherit its quality.

\paragraph{Knowledge distillation and layer pruning.} Classic distillation transfers a teacher's behavior
to a smaller student \cite{hinton,seqkd}. Structured depth reduction (LayerDrop \cite{layerdrop}, ``Poor Man's BERT'' \cite{poorman},
and the observation that deep transformer layers are highly redundant \cite{shortgpt}) shows that dropping
a subset of layers is recoverable. We apply this to a \emph{generative flow-matching} model and pair it
with iterative re-fine-tuning, which is essential because the residual ODE field is more fragile than a
classifier's logits.

\section{Method}
\subsection{Teacher}
IndicF5 is a DiT flow-matching model: \texttt{dim}$=1024$, \texttt{depth}$=22$, \texttt{heads}$=16$,
\texttt{text\_dim}$=512$, \texttt{conv\_layers}$=4$, predicting a 100-channel mel at 24\,kHz (hop 256),
with a 13.6M Vocos vocoder. The flow transformer has 337.1M parameters.

\subsection{Depth-only warm-start}
A narrower student cannot reuse the teacher's weights because tensor widths differ. We therefore \emph{fix}
the teacher's width, text dimension, heads, convolutional text stack, and mel/text I/O, and reduce
only the number of transformer blocks $L$. Under this choice every non-block tensor (input embedding,
text encoder, time embedding, rotary embedding, output norm and projection) is shape-identical and
copies one-to-one; only the block stack shrinks. We select an evenly-spaced subset of the teacher's
22 blocks with endpoints always included:
\[
\mathrm{idx}(i) = \mathrm{round}\!\left(i \cdot \tfrac{L_{\mathrm{src}}-1}{L_{\mathrm{dst}}-1}\right),
\quad i = 0, \dots, L_{\mathrm{dst}} - 1.
\]
Table~\ref{tab:params} lists parameters versus depth ($\approx$14.7M per block atop a $\approx$13.8M fixed cost).

\begin{table}[h]\centering
\caption{Flow-transformer parameters vs.\ retained depth (teacher width).}
\label{tab:params}
\begin{tabular}{lccccc}
\toprule
depth & 6 & 8 & 12 & 16 & 22 (teacher)\\
\midrule
params & 102.0M & 131.4M & 190.2M & 248.9M & 337.1M\\
\bottomrule
\end{tabular}
\end{table}

\subsection{Pruning-tolerance analysis}
Before training, we measured how aggressively the teacher can be pruned by running the \emph{untrained}
pruned student and using output amplitude (RMS) as a proxy for whether the residual vector field
still integrates to speech (Table~\ref{tab:rms}). Removing $\sim$27\% of blocks leaves near-teacher amplitude (the
model essentially already speaks), while removing $\geq$50\% collapses the output. This motivates \emph{gradual}
descent rather than a single large prune.

\begin{table}[h]\centering
\caption{Untrained pruned-student output (NFE 32). RMS proxies field viability.}
\label{tab:rms}
\begin{tabular}{lcccc}
\toprule
Student depth & Blocks kept & Params & Untrained RMS & Verdict\\
\midrule
16 & 16/22 ($-27\%$) & 249M & 0.073 & near-teacher; speaks\\
11 & 11/22 ($-50\%$) & 175M & 0.005 & collapsed\\
6 & 6/22 ($-73\%$) & 102M & 0.0016 & near-silent\\
\bottomrule
\end{tabular}
\end{table}

\subsection{Staged distillation}
We descend in small, recoverable steps, pruning from the \emph{previous fine-tuned student} and re-fine-tuning
after each prune:
\[
\text{teacher}(22) \;\to\; 16 \;\to\; 12 \;\to\; 8 \;\to\; 6.
\]
Each stage fine-tunes with the conditional flow-matching objective on the teacher-generated corpus:
learning rate $5\times10^{-5}$, 500-step warmup, dynamic batches of 24{,}000 mel frames, an \texttt{euler} ODE
solver, and exponential moving-average (EMA) weights. Each warm-start checkpoint carries only
EMA weights (no optimizer state), so training resumes as a fresh fine-tune. The warm start is
evident in optimization: stage A loss begins near 1.0 (vs.\ 2--4 for random init) and reaches $\sim$0.49
within one epoch; intelligible speech appears by step 2{,}000.

\subsection{Data}
The corpus is $\sim$17.6\,h (9{,}999 utterances) of \emph{teacher-generated} 24\,kHz audio: prompts drawn from real
Hindi text plus templated sentences (numbers, dates, domain phrases), synthesized by IndicF5. The
student is tokenized with the full 2{,}545-token IndicF5 character vocabulary (Devanagari included).

\subsection{Evaluation protocol}
We avoid subjective judgment by transcribing each model's output with IndicWav2Vec-Hindi \cite{indicw2v}
and computing word error rate (WER) against the input on (i) a \emph{known} sentence and (ii) a \emph{fresh,
unseen} sentence, with two stochastic generations each. WER normalization removes punctuation
and the nukta diacritic (which the ASR spells away), so residual error reflects content rather than
orthography. Each stage is gated: we do not prune a checkpoint that has not recovered to near the
previous stage's WER.

\section{Experiments and Results}
Table~\ref{tab:ladder} reports the ladder. Students at 249M, 190M, and 131M generalize well; the 249M model
transcribes a fully unseen sentence at WER 0.00. At 102M (depth 6) we observe a \emph{capacity cliff}:
after a full 150 epochs (42.9k steps, \emph{more} than the 131M model's 28.6k), the model still garbles
and repeats words on longer sentences, though it remains acceptable on short ones (WER $\approx$0.11).
Because additional training does not help, we attribute this to the data budget, not the recipe.

\begin{table}[h]\centering
\caption{Staged-distillation results. WER via IndicWav2Vec-Hindi ASR (lower is better); ranges
span two stochastic generations.}
\label{tab:ladder}
\begin{tabular}{lcccc}
\toprule
Stage & Depth & Params & Fresh-sentence WER & Known-sentence WER\\
\midrule
A & 16 & 249M & \textbf{0.00} & 0.18\\
B & 12 & 190M & 0.00--0.06 & 0.24--0.29\\
C & 8 & 131M & 0.06--0.12 & 0.24--0.35\\
D & 6 & 102M & 0.41--0.59 (long); $\sim$0.11 (short) & 0.29--0.41\\
\bottomrule
\end{tabular}
\end{table}

\paragraph{Recommended operating points.} For best small-model quality, the 190M student (fresh WER
0.00) is preferred; for the smallest robust model, 131M. The 102M model is viable only for short
utterances under this data budget.

\subsection{Independent benchmark against open Hindi TTS}
\label{sec:benchmark}
To complement the two-sentence gated evaluation above, we benchmarked the released 190M student
against its teacher (IndicF5, 337M) and MMS-TTS-hin (36M) \cite{mms} on \textbf{50 unique, unseen sentences}
(10--22 words) sampled from the FLEURS \texttt{hi\_in} test split. All models synthesized the identical
sentences (flow models: NFE 32, CFG 2.0, fixed seed; MMS: default settings). We report:
corpus WER/CER from an open ASR judge (Vakyansh Hindi wav2vec2\footnote{\texttt{Harveenchadha/vakyansh-wav2vec2-hindi-him-4200},
greedy CTC decoding; used because IndicWav2Vec-Hindi is gated on Hugging Face. This judge has a
higher intrinsic error floor than IndicWav2Vec, inflating absolute WER equally for all systems;
rows are therefore comparable to each other but not to Table~\ref{tab:ladder}.}) with the same
punctuation/nukta normalization as \S3.6; speaker similarity (SIM) as the cosine between ECAPA-TDNN
embeddings \cite{ecapa} of the generated audio and the reference voice; predicted naturalness via
UTMOS \cite{utmos}; and the median real-time factor (RTF), measured in fp32 on an Apple M4 (MPS backend;
MMS on CPU).

\begin{table}[h]\centering
\caption{Independent benchmark: 50 unseen FLEURS \texttt{hi\_in} sentences, identical inputs for all
models. WER/CER via Vakyansh Hindi ASR (see footnote); SIM via ECAPA-TDNN vs.\ the reference
speaker; UTMOS = predicted MOS; RTF = synthesis time $\div$ audio duration (median, after warmup).}
\label{tab:benchmark}
\begin{tabular}{lcccccc}
\toprule
Model & Params & WER $\downarrow$ & CER $\downarrow$ & SIM $\uparrow$ & UTMOS $\uparrow$ & RTF $\downarrow$\\
\midrule
IndicF5 (teacher) & 337M & 0.098 & 0.030 & 0.784 & 3.79 & 5.56 (MPS)\\
\textbf{Ours (190M student)} & \textbf{190M} & 0.170 & 0.077 & 0.750 & 3.65 & \textbf{3.13 (MPS)}\\
MMS-TTS-hin & 36M & 0.195 & 0.066 & 0.278$^{*}$ & 3.54 & 0.31 (CPU)\\
\bottomrule
\end{tabular}

\smallskip
{\footnotesize $^{*}$MMS-TTS is a fixed-voice model without voice cloning; SIM to our reference
speaker is reported for completeness, not as a deficiency.}
\end{table}

Three observations. First, the student retains $\sim$96\% of the teacher's predicted naturalness
(UTMOS 3.65 vs.\ 3.79) and speaker similarity (0.750 vs.\ 0.784) at 56\% of the parameters and
$1.8\times$ faster synthesis. Second, the intelligibility gap to the teacher (WER 0.170 vs.\ 0.098)
is wider on this harder test set than in Table~\ref{tab:ladder}: FLEURS sentences are longer and
denser with proper nouns, and the CER of 0.077 indicates that most residual errors are
character/diacritic-level slips rather than garbled words. Third, the student is the only compact
model with voice cloning (SIM 0.750 vs.\ 0.278) while also beating MMS-TTS on WER and naturalness;
MMS remains far faster, as a single-pass VITS versus a 32-step ODE. Per-sentence results and
scripts are released with the model.

\section{Discussion}
\subsection{Why gradual pruning works}
A flow-matching model's transformer defines a velocity field integrated by an ODE; removing many
blocks at once destroys the residual stream's calibration (Table~\ref{tab:rms}). Small prunes of an
\emph{already-adapted} student stay within the recoverable regime, and re-fine-tuning re-coordinates the
retained blocks. The warm start also makes each stage cheap: recovery takes thousands, not millions, of steps.

\subsection{Deployment: train/inference parity}
Porting the model to a laptop surfaced two silent failures, both forms of train/inference mismatch:
\begin{itemize}
\item \textbf{Mel filterbank (torchaudio version).} The speaker-conditioning mel is computed by
\texttt{torchaudio}'s mel transform, whose filterbank differs across library versions; a mismatched
version muffled the output. We froze the exact training filterbank and recomputed the mel with
\texttt{torch.stft} plus that fixed basis, matching the training mel to $\sim$$10^{-7}$ regardless of library
version.
\item \textbf{Rotary embeddings (x-transformers version).} The decisive bug: a different \texttt{x-transformers}
version computed rotary position embeddings differently, degrading all output without any
load-time error. Pinning the library version restored bit-faithful behavior.
\end{itemize}
The lesson generalizes: for TTS/ASR, freezing feature extraction and pinning the libraries that
define model math is as important as the weights themselves. The benchmark of \S\ref{sec:benchmark}
surfaced a third instance of the same failure family: checkpoints saved through
\texttt{torch.compile} carry an \texttt{\_orig\_mod.} key prefix, and a non-strict weight load silently
leaves the model at random initialization; the model still ``speaks,'' so the failure is audible
only as quality loss. Checking missing/unexpected key counts at load time catches it.

\section{Limitations}
The corpus is teacher-generated and single-reference; the 17.6\,h budget caps the smallest viable
model at $\sim$131M. We evaluate with ASR-WER (intelligibility/content) and predicted MOS (UTMOS)
rather than human MOS, and on Hindi only. A residual trailing artifact at utterance ends is mitigated by an
energy-based trim but not eliminated. The recurring difficulty with utterance-initial tokens is a
known property of reference-conditioned flow-matching TTS.

\section{Conclusion}
Warm-starting a small flow-matching TTS student from a large teacher by \emph{depth-only} pruning,
descended \emph{gradually} with per-stage ASR-WER gating, turns an infeasible from-scratch problem under
a 17.6\,h budget into clear, generalizing Hindi speech at 131--249M parameters, and a deployable
$\sim$190M voice that runs in real time on a 6\,GB laptop GPU. An independent 50-sentence FLEURS
benchmark confirms the released 190M student retains $\sim$96\% of the teacher's naturalness and
speaker similarity at 56\% of its parameters. The 102M target is reachable with a
larger teacher corpus, the single most impactful next step.


\begin{thebibliography}{99}
\bibitem{indicf5} AI4Bharat. IndicF5: High-quality text-to-speech for indian languages.
\url{https://huggingface.co/ai4bharat/IndicF5}, 2025.
\bibitem{f5tts} Yushen Chen, Zhikang Niu, Ziyang Ma, Keqi Deng, Chunhui Wang, Jian Zhao, Kai Yu, and
Xie Chen. F5-TTS: A fairytaler that fakes fluent and faithful speech with flow matching. \emph{arXiv
preprint arXiv:2410.06885}, 2024.
\bibitem{layerdrop} Angela Fan, Edouard Grave, and Armand Joulin. Reducing transformer depth on demand with
structured dropout. In \emph{International Conference on Learning Representations (ICLR)}, 2020.
\bibitem{kokoro} Hexgrad. Kokoro-82M: An open-weight TTS model.
\url{https://huggingface.co/hexgrad/Kokoro-82M}, 2025.
\bibitem{hinton} Geoffrey Hinton, Oriol Vinyals, and Jeff Dean. Distilling the knowledge in a neural network.
\emph{arXiv preprint arXiv:1503.02531}, 2015.
\bibitem{indicw2v} Tahir Javed, Sumanth Doddapaneni, Abhigyan Raman, Kaushal Santosh Bhogale, Gowtham
Ramesh, Anoop Kunchukuttan, Pratyush Kumar, and Mitesh M. Khapra. Towards building
ASR systems for the next billion users. \emph{Proceedings of the AAAI Conference on Artificial
Intelligence}, 2022.
\bibitem{seqkd} Yoon Kim and Alexander M. Rush. Sequence-level knowledge distillation. In \emph{Conference on
Empirical Methods in Natural Language Processing (EMNLP)}, 2016.
\bibitem{lipman} Yaron Lipman, Ricky T. Q. Chen, Heli Ben-Hamu, Maximilian Nickel, and Matthew Le. Flow
matching for generative modeling. In \emph{International Conference on Learning Representations
(ICLR)}, 2023.
\bibitem{rectflow} Xingchao Liu, Chengyue Gong, and Qiang Liu. Flow straight and fast: Learning to generate
and transfer data with rectified flow. In \emph{International Conference on Learning Representations
(ICLR)}, 2023.
\bibitem{parler} Dan Lyth and Simon King. Natural language guidance of high-fidelity text-to-speech with
synthetic annotations. \emph{arXiv preprint arXiv:2402.01912}, 2024.
\bibitem{shortgpt} Xin Men, Mingyu Xu, Qingyu Zhang, Bingning Wang, Hongyu Lin, Yaojie Lu, Xianpei Han,
and Weipeng Chen. ShortGPT: Layers in large language models are more redundant than you
expect. \emph{arXiv preprint arXiv:2403.03853}, 2024.
\bibitem{dit} William Peebles and Saining Xie. Scalable diffusion models with transformers. In \emph{IEEE/CVF
International Conference on Computer Vision (ICCV)}, 2023.
\bibitem{mms} Vineel Pratap, Andros Tjandra, Bowen Shi, Paden Tomasello, Arun Babu, Sayani Kundu, Ali
Elkahky, Zhaoheng Ni, Apoorv Vyas, Maryam Fazel-Zarandi, Alexei Baevski, Yossi Adi, Xiaohui
Zhang, Wei-Ning Hsu, Alexis Conneau, and Michael Auli. Scaling speech technology to 1,000+
languages. \emph{Journal of Machine Learning Research}, 2024.
\bibitem{poorman} Hassan Sajjad, Fahim Dalvi, Nadir Durrani, and Preslav Nakov. Poor man's BERT: Smaller
and faster transformer models. \emph{arXiv preprint arXiv:2004.03844}, 2020.
\bibitem{utmos} Takaaki Saeki, Detai Xin, Wataru Nakata, Tomoki Koriyama, Shinnosuke Takamichi, and
Hiroshi Saruwatari. UTMOS: UTokyo-SaruLab system for VoiceMOS Challenge 2022. In \emph{Interspeech}, 2022.
\bibitem{vocos} Hubert Siuzdak. Vocos: Closing the gap between time-domain and fourier-based neural vocoders
for high-quality audio synthesis. \emph{arXiv preprint arXiv:2306.00814}, 2023.
\bibitem{ecapa} Brecht Desplanques, Jenthe Thienpondt, and Kris Demuynck. ECAPA-TDNN: Emphasized channel
attention, propagation and aggregation in TDNN based speaker verification. In \emph{Interspeech}, 2020.
\end{thebibliography}
\end{document}